\documentclass[twocolumn,multicolumn,aps,prb,showpacs]{revtex4}

\usepackage{graphicx}
\usepackage{dcolumn}
\usepackage{bm}
\usepackage{color}
\usepackage{latexsym}
\input{epsf}

\begin{document}

\preprint{APS/123-QED}

\title{ Thermodynamics of the vortex liquid in heavy ion-irradiated 
superconductors}
%
%
%
\author{Cornelis J. van der Beek, Marcin Konczykowski}
\affiliation{
Laboratoire des Solides Irradi\'{e}s, CNRS-UMR 7642 \&
CEA/DSM/DRECAM,
Ecole Polytechnique, 91128 Palaiseau cedex, France}
\author{Luc Fruchter}
\affiliation{%
Laboratoire de Physique des Solides, B\^{a}timent 510,
Universit\'{e} Paris-Sud, CNRS, 91405 Orsay, France
}%
\author{Ren\'{e} Brusetti, Thierry Klein, and Jacques Marcus}
\affiliation{
Laboratoire d'Etudes des Propri\'{e}t\'{e}s Electroniques des
Solides, Centre National de la Recherche Scientifique,  B.P. 166, 38042
Grenoble cedex 9, France
}%
\author{Christophe Marcenat}
\affiliation{D\'{e}partement de Recherche Fondamentale sur la
Mati\`{e}re Condens\'{e}e, Service de Physique des Solides
Magn\'{e}tiques et Supraconducteurs, Commissariat \`{a} l'Energie 
Atomique, 17 Avenue
des Martyrs, 38054, Grenoble cedex 9, France}

\date{\today}

\begin{abstract}
It is shown that the large effect of heavy ion-irradiation on the thermodynamical properties of the
anisotropic superconductor YBa$_{2}$Cu$_{3}$O$_{7-\delta}$ extends
well into the superconducting fluctuation regime. The presence of the induced amorphous
columnar defects shifts the specific heat maximum at the normal-to-superconducting transition.
This effect is similar to that recently put into evidence in cubic
K$_{x}$Ba$_{1-x}$BiO$_{3}$ ($x \simeq 0.35$). In both compounds, vortex pinning
manifests itself as a sharp angular dependence of the \em equilibrium \rm
torque. In YBa$_{2}$Cu$_{3}$O$_{7-\delta}$, pinning by the defects 
appears at the
temperature $T_{C_{p}}^{max}$ of the specific heat maximum, well above the
magnetic irreversibility line $T_{irr}(H)$. In isotropic 
K$_{x}$Ba$_{1-x}$BiO$_{3}$,
the onset of the pinning-related torque anomaly tracks the onset of
the specific heat anomaly and the irreversibility line. In 
YBa$_{2}$Cu$_{3}$O$_{7-\delta}$,
fluctuations of the amplitude of the order parameter (and not vortex 
line wandering)
are ultimately responsible for the vanishing of pinning. In 
K$_{x}$Ba$_{1-x}$BiO$_{3}$,
vortex pinning disappears only at the superconducting-to-normal transition.
The results indicate that in both compounds, the pinning energy at 
the ``Bose glass'' transition is large with 
respect to the total
free energy gain in the superconducting state. By implication, the 
mechanism of this latter transition should be reconsidered.
\end{abstract}

\maketitle

\section{Introduction}

When heated above their irreversibility line, disordered
type II superconductors undergo a transition from a ``truly superconducting''
ensemble of localized vortex lines to a ``vortex liquid'' of diffusing lines.
This transition is most commonly described in terms of a thermal 
``depinning'' of
vortex lines from material defects. Typically, thermal
wandering of the vortex lines from the defects is thought to become
increasingly important as the temperature is raised, until, at the
transition, the free energy gain obtained from vortex localization on
the defects has dropped to $\sim k_{B}T$.

The case where the defects are columnar amorphous tracks introduced 
by swift heavy-ion
irradiation has attracted much attention, not in the least because the problem
becomes particularly tractable theoretically. Using the formal analogy between
flux lines and 2D bosons in a static disorder potential, Nelson and
Vinokur \cite{nelson93} calculated the magnitude of
thermal positional excursions, concomitant pinning energies, and
the resultant phase diagram. This consists of a low temperature
disordered ``Bose glass'' of localized vortices that gives way to the
vortex liquid at the temperature $T_{BG}(B)$ [or induction $B_{BG}(T)$].
Model descriptions of the Bose glass-to-liquid transition nearly 
exclusively rely on
the total pinning energy near $T_{BG}(B)$ being small. This can be 
either because
line wandering reduces the single-vortex pinning 
energy,\cite{nelson93,Samoilov96}
or because vortices vastly outnumber the defects. In
both cases, the pinning energy gain is only a small perturbation to the
total free energy of the system, and the position of the Bose-glass 
transition line
can be obtained as a 
shift\cite{nelson93,Samoilov96,Blatter94,KrusinElbaum94,Larkin95,Blatter2002}
of the (first order) vortex lattice melting line of the pristine
material.\cite{Safar92II,Kwok92,Schilling96}
This approach has had some success in explaining the observed 
increase of $B_{BG}$ with defect
density $n_{d}$ in YBa$_{2}$Cu$_{3}$O$_{7-\delta}$, where it was found
that $\partial B_{BG}/\partial T \sim 1 + A \Phi_{0}^{1/2} n^{1/2}_{d}$
(with $\Phi_{0} = h/2e$ the flux quantum).\cite{KrusinElbaum94,Samoilov96}

Simultaneously, experiments abound indicating that the pinning energy near
the Bose-glass transition in cuprate superconductors is {\em not} 
small. Reversible
magnetization measurements on heavy-ion irradiated layered 
Bi$_{2}$Sr$_{2}$CaCu$_{2}$O$_{8}$
reveal a large contribution of columnar defect pinning to the free
energy.\cite{vdBeek96,vdBeek2000} An effect of the ion tracks was measured up
into the fluctuation critical regime\cite{vdBeek96}, and interpreted 
in terms of local
$T_{c}$--variations induced by the defects.\cite{vdBeek96,Braverman2002}
Recent measurements on cubic (K,Ba)BiO$_{3}$ have shown that heavy-ion
irradiation even affects the specific heat $C_{p}$: the temperature 
at which the $C_{p}$ jump occurs, signaling the transition to the 
superconducting state,
was found to shift upward with increasing defect density $n_{d}$, and 
depends on the
angle between magnetic field and the track 
direction.\cite{Marcenat2003,Klein2004}
When the magnetic field is aligned with the tracks, superconductivity 
is enhanced,
when the magnetic field is turned away from the tracks, one recovers 
the behavior
of the pristine crystal. As for heavy-ion irradiated YBa$_{2}$Cu$_{3}$O$_{7-\delta}$, 
transport experiments show that in the vortex liquid, the resistivity 
remains exponentially
small with respect to that of the pristine 
material.\cite{Jiang94,Paulius97,Kim98,Kwok98} More
strikingly, the experiments of 
Refs.~\onlinecite{Paulius97,Kim98,Kwok98} reveal an angular
dependence of the resistivity, related to vortex pinning
by the tracks, that persists up to resistance levels that are 90
percent of that in the normal state. In other words, columnar
defects affect vortices in the vortex liquid, and up into the
fluctuation paraconductivity regime. However, there are no
thermodynamic measurements assessing the importance of the pinning
energy in this material.


In order to  establish the magnitude of the contribution
of vortex pinning by columnar defects to the free energy, we have 
performed measurements
of the specific heat and the reversible torque on heavy ion-irradiated single
crystalline YBa$_{2}$Cu$_{3}$O$_{7-\delta}$ (Sections~\ref{section:Cp}
and \ref{torque}). The data are compared to previous
specific heat results\cite{Marcenat2003,Klein2004} and
new torque measurements on single crystalline
K$_{x}$Ba$_{1-x}$BiO$_{3}$. The main difference between the two
materials lies in their Ginzburg number $Gi \equiv \frac{1}{2} \left[
k_{B}T_{c}/2\pi\varepsilon\varepsilon_{0}(0)\xi(0) \right]^{2}$. Here
$\varepsilon_{0}(T) = \Phi_{0}^{2}/4\pi\mu_{0}\lambda_{ab}^{2}(T)$ is 
the vortex energy scale,
$\lambda_{ab}(T)$ is the penetration depth for currents running 
perpendicularly to the
crystalline anisotropy axis, and $\xi(T)$ the coherence
length (for YBa$_{2}$Cu$_{3}$O$_{7-\delta}$, $\xi = \xi_{ab}$, the 
$ab$-plane coherence
length). Since $Gi$ for optimally doped YBa$_{2}$Cu$_{3}$O$_{7-\delta}$
is two orders of magnitude larger than that of K$_{x}$Ba$_{1-x}$BiO$_{3}$
(see Table~\ref{tbl:parameters}), it is conceivable that thermal fluctuations
wipe out any strong effect of columnar defects in the vortex liquid phase.
On the contrary, we find that the reduced temperature
$T_{C_{p}}^{max}(H)/T_{c}$ at which the specific heat is maximum in
YBa$_{2}$Cu$_{3}$O$_{7-\delta}$ unambiguously \em shifts upward \rm with
increasing columnar defect density $n_{d}$, as it does
in K$_{x}$Ba$_{1-x}$BiO$_{3}$. This upward shift can only be
accounted for by a large (\em i.e. \rm not perturbatively small)
contribution of pinning by the columnar defects to the
free energy. This is in contradiction to the assumptions commonly made
in estimating the Bose-glass transition line.
\cite{nelson93,Blatter94,KrusinElbaum94,Larkin95,Samoilov96,Blatter2002}
In Section~\ref{section:suppression},
we estimate the mean-field pinning energy contribution required for the
specific-heat shift.

We also obtain the experimental pinning energy directly
from reversible torque measurements (Section \ref{section:pinning}).
It turns out that the field- and temperature dependence of the
pinning energy, rather surprisingly, scales with the parameter
$Q = (1-b) (1-t^{2})^{1/3} (tb)^{-2/3} Gi^{-1/3}$, suggesting that a 
development of
the Ginzburg-Landau free energy functional in terms of Lowest Landau 
level (LLL)
eigenfunctions is an appropriate starting point for a model
description.\cite{Ikeda89,Welp91,Sasik95,DingpingLi2001,DingpingLi2002,Mikitik2003} 

Here, $t \equiv T/T_{c}^{MF}$ with $T_{c}^{MF}$ the mean-field transition
temperature, and $b \equiv B/B_{c2}(T)$ with $B_{c2}(T) =
\Phi_{0}/2\pi\xi^{2}(T)$ the upper critical field.
An assessment of the experimental result shows that in 
YBa$_{2}$Cu$_{3}$O$_{7-\delta}$,
fluctuations of the order parameter \em amplitude \rm lower the pinning energy
with respect to the expected mean-field value. In
K$_{x}$Ba$_{1-x}$BiO$_{3}$, with small $Gi \sim 10^{-5}$, thermal
fluctuations are unimportant and pinning subsists up to the
superconducting-to-normal state transition.

\section{Experiments and Results}
\label{results}

\subsection{Samples}

Experiments were done on a series of untwinned and lightly twinned
YBa$_{2}$Cu$_{3}$O$_{7-\delta}$ single crystals, grown by the flux
method in Au crucibles, and subsequently annealed in oxygen in Pt
tubes.\cite{Holtzberg} Measurements on K$_{x}$Ba$_{1-x}$BiO$_{3}$
were made on a crystal with $x = 0.35$, grown by electrocrystallization.
The crystals were irradiated during different runs at the Grand
Acc\'{e}l\'{e}rateur National d'Ions Lourds (GANIL) in Caen, France.
The YBa$_{2}$Cu$_{3}$O$_{7-\delta}$ crystals were irradiated with 5.8
GeV Pb ions; for all but one crystal, the beam was aligned  parallel to 
the $c$-axis. The final crystal (hereafter referred to as ``Y30'')  
was at an angle of $30^{\circ}$ with respect to the $c$ axis, to a fluence of
$1 \times 10^{11}$ ions cm$^{-2}$, which corresponds to a matching field
$B_{\phi} \equiv \Phi_{0}n_{d}= 2$ T.  The K$_{0.35}$Ba$_{0.65}$BiO$_{3}$ 
crystal was irradiated with 7.2 GeV Ta ions. The irradiation produced continuous 
amorphous columnar defects of radius $c_{0} \approx 3.5$ nm. The irreversibility 
line $T_{irr}(H)$

\begin{table}[b]
     \centering
     \begin{tabular}{ccc}
	\hline
   	\hline & & \\
	       & YBa$_{2}$Cu$_{3}$O$_{7-\delta}$ & 
K$_{0.35}$Ba$_{0.65}$BiO$_{3}$ \\
	       & & \\
	\hline
	\hline
	& & \\
	$\lambda$            & $\lambda_{ab}(0) =120$ nm      & 
$\lambda(0) = 220$ nm  \\
	& & \\
	$\xi$                & $\xi_{ab}(0) =  1.4$ nm        & 
$\xi(0) = 3.8$ nm             \\
	& & \\
	$\varepsilon_{0}(0)$ & $1.7\times 10^{-11}$ Jm$^{-1}$ & 
$5.2\times 10^{-11}$ Jm$^{-1}$\\
	& & \\
	$Gi$                 &      $2\times10^{-3}$          & 
$1\times10^{-5} $ \\
	& & \\
	$\varepsilon = \lambda_{ab}/\lambda_{c} = \xi_{c}/\xi_{ab}$ & 
0.14 & 1 \\
	& & \\
	\hline
	\hline
\end{tabular}
     \caption{Superconducting parameters for the studied compounds.}
     \label{tbl:parameters}
\end{table}

\noindent of all crystals was measured
as the onset temperature of the third harmonic of the ac
transmittivity $T_{H3}$,\cite{Konczykowski93,vdBeek95} with the DC field
aligned parallel to the defect direction. The irreversibility line of
the ``Y30'' and K$_{0.35}$Ba$_{0.65}$BiO$_{3}$ crystals was also obtained
from torque magnetometry (see Table~\ref{tbl:crystals}).

\subsection{Torque measurements}
\label{torque}

Apart from the characterization of the pinning energy,
torque measurements were also used
to obtain the superconducting parameters of the two compounds under
study (see Table~\ref{tbl:parameters}).
The measurements on YBa$_{2}$Cu$_{3}$O$_{7-\delta}$ were
performed on the same twinned single crystal (Y30) as that used in
Ref.~\onlinecite{hayani2000}. 
The microtorque setup\cite{hayani2000} was 
improved by the adjunction of a secondary

\begin{figure}[t]
\centerline{\epsfxsize 6cm \epsfbox{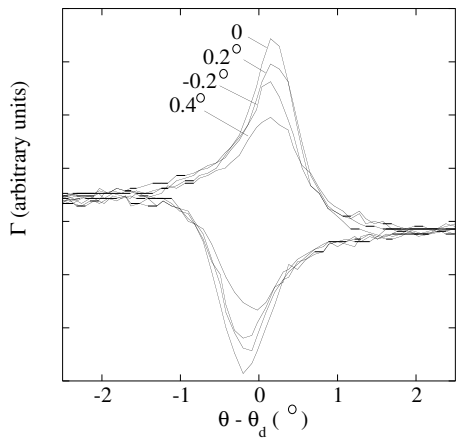}}
\caption{YBa$_{2}$Cu$_{3}$O$_{7-\delta}$ : Torque signal for rotation
angles   $\theta$ close to the irradiation direction $\theta_{d}$, 
for different angles
$\vartheta$ between the plane of rotation of the magnetic field and 
the defects ($H =
20$ kOe, $T = 88$ K).}
\label{alpha}
\end{figure}

\begin{table}[b]
     \centering
     \begin{tabular}{cccccccc}
	\hline
	\hline
	& & & & & & & \\
	& YBa$_{2}$Cu$_{3}$O$_{7-\delta}$ & & 
YBa$_{2}$Cu$_{3}$O$_{7-\delta}$& &
YBa$_{2}$Cu$_{3}$O$_{7-\delta}$ 	 & K$_{0.35}$Ba$_{0.65}$BiO$_{3}$ \\

	& pristine & $B_{\phi} = 1$ T & $B_{\phi} = 2$ T & $B_{\phi} = 5$ T
& $B_{\phi} = 2$ T, 30$^{\circ}$ & $B_{\phi} = 2$ T    \\
& & & & & & &\\
	  \hline
	  \hline
	&                 &               &
&                         &                         &
&  \\
Dimensions ($\mu$m$^{3}$) &$330\times 400$&  triangle of base
650,      &$430\times 510$ & $200\times 450$&
$130\times 337\times 18$ & $120 \times 45 \times 30$ & \\
			  &  $\times 20$  &   height  530, thickness 20
&   $\times 20$      & $\times 20$ & ``Y30''&\\
			  &                 &               &
&                         &                         &
&  \\
Description	          & untwinned     & 16 TB's $\parallel$ base &
TB's spaced   &  twinned  & TB's spaced & tracks $\parallel$  \\
			  & rectangle     & 
& by 10 $\mu$m
&           &  by 5 $\mu$m &long dimension  \\
			     &      &      &      &      &   &       \\
$T_{C_{p}}^{max}(H=0)$ ( K ) & 93.1 & 92.4 & 92.1 & 91.0 & * & *     \\

$T_{irr}(H=0)$ ( K )      & 93.1 & 92.5 & 92.0 & 90.8 & 91.3 & 31.5     \\
$T_{c}^{MF}$ ( K )        & 93.1 & 93.1 & 93.1 & 93.1 & 92.3 & 32.4     \\

	 &   &   &   &   &   &   &    \\
	\hline
	\hline
     \end{tabular}
     \caption{Characteristics of
     single crystals used in this study. All crystals were either
     untwinned or twinned with a single twin boundary (TB) orientation.}
     \label{tbl:crystals}
\end{table}
%
%
%
%
\noindent magnetic field perpendicular to the main one, so that the plane in
which the field is rotated could be chosen arbitrarily. The angle $\vartheta$
between this plane and the defect direction could be set with a resolution
better than $0.1^{\circ}$. We set $\vartheta$ to zero $\pm 0.1^{\circ}$ by
maximizing the pinning measured by the torque irreversibility along the defect
direction (Fig.~\ref{alpha}).

\begin{figure}[t]
\centerline{\epsfxsize 8.5cm \epsfbox{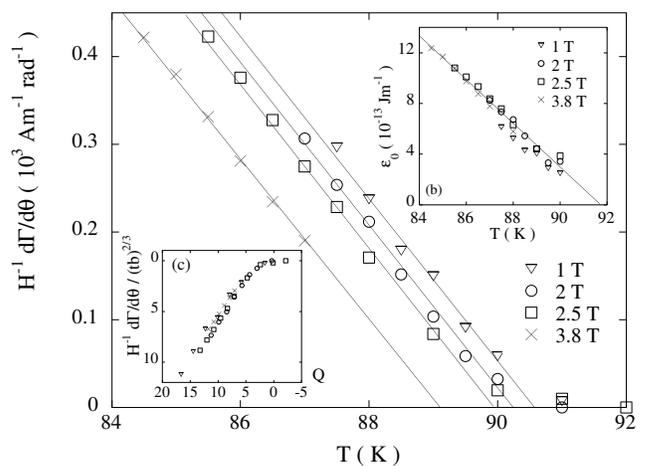}}
  \caption{YBa$_{2}$Cu$_{3}$O$_{7-\delta}$ : Slope $\partial \Gamma /
	  \partial \theta$ of the torque signal for field along the $c$
	  axis, divided by field.  Lines are linear fits extrapolating to $T_{c}(H)$.
	   Inset (b): line energy $\varepsilon_{0}(T)$ obtained by dividing out the field dependence
	   from the data from the data in the main panel. Inset (c): Scaled 
	   magnetization $H^{-1}d\Gamma/d\theta / (tb)^{2/3}$ as function of the 
           LLL parameter $Q = (1-b)(1-t^{2})^{1/3}/(tb)^{2/3}Gi^{1/3}$.}
\label{Hc2}
\vspace{-1mm}
\end{figure}

The mean field upper critical field line $B_{c2}(T)$ was located using the slope of the equilibrium
torque per unit volume. For small angles
$\theta$ between the applied field and the $c$ axis direction, this 
is given  by
\begin{equation}
d\Gamma/d\theta \simeq H M_{\bot}(H) ,
\end{equation}
where $M_{\bot}$ is the magnetization for the
field applied along the $c$--axis.\cite{lebras96,hao91,buzdin94}
A plot of $H^{-1}d\Gamma/d\theta$, shown in Fig.~\ref{Hc2}, 
represents $M_{\bot}(H)$. For the lower temperatures,  
$H^{-1}d\Gamma/d\theta$  depends linearly on $T$, in agreement
with the (mean-field)

  \newpage
  \noindent  Abrikosov formula 
\begin{equation}
     - M_{\bot} \approx \frac{\varepsilon_{0}}{\beta_{A} \Phi_{0}}
     (1-b),  \hspace{1cm}(b\lesssim 1)
     \label{eq:abrikosov}
\end{equation}

\begin{figure}[t]
\centerline{\epsfxsize 8cm \epsfbox{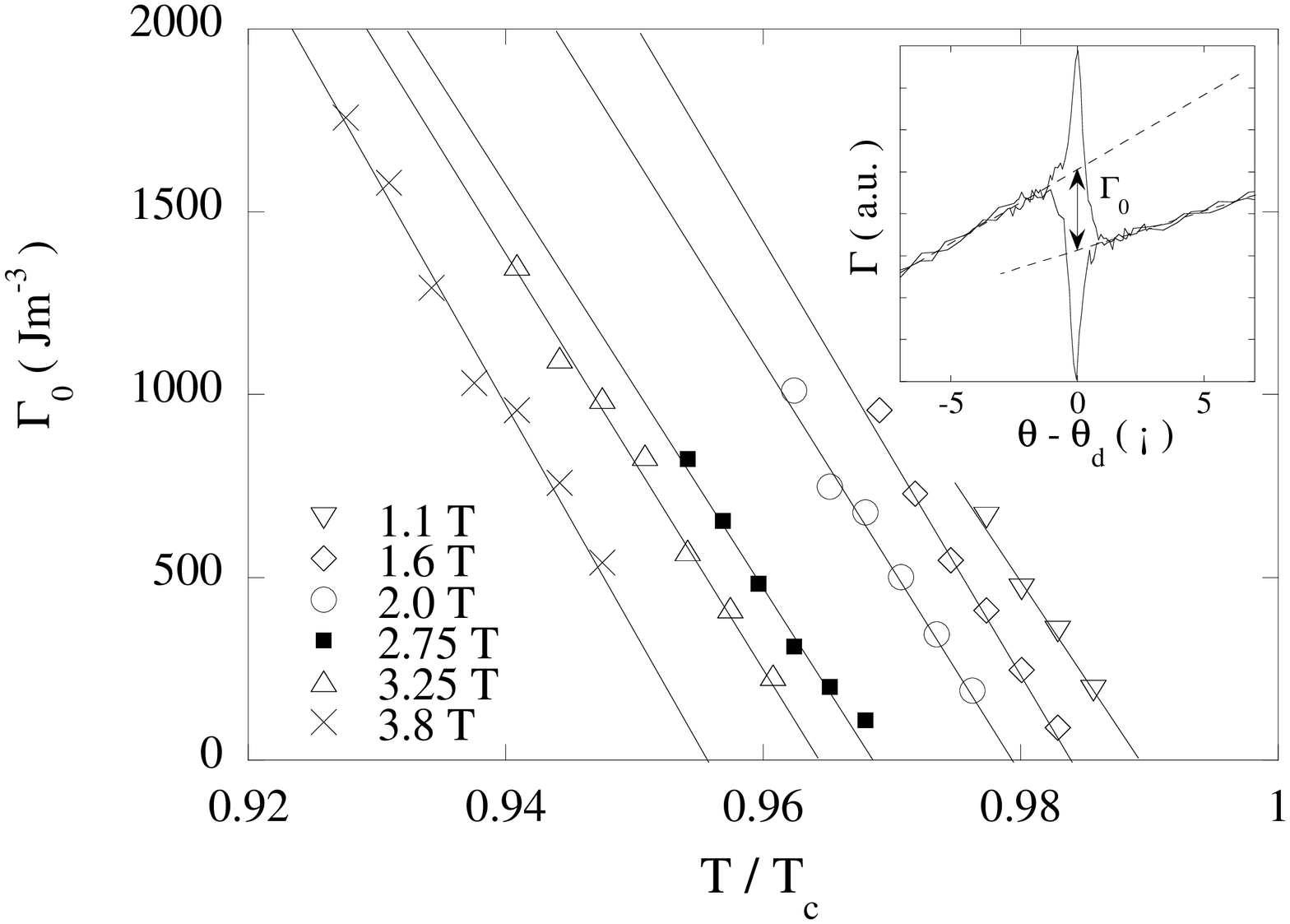}}
  \caption{YBa$_{2}$Cu$_{3}$O$_{7-\delta}$: magnitude of the torque
  discontinuity at field alignment with the defect direction $\theta_{d}$.
  Straight lines represent the extrapolation of the torque discontinuity
  to $T_{k}(H)$.  The inset shows the torque curves for both directions
  of rotation at $\mu_{0}H= 2$ T and $T = 0.97 T_{c}$. It also shows
  the determination of the torque discontinuity $\Gamma_{0}$ by
  extrapolation of the linear signal to $\theta = 0$.  }
\label{step}
\end{figure}

\begin{figure}[b]
    \vspace{-0mm}
\centerline{\epsfxsize 8cm \epsfbox{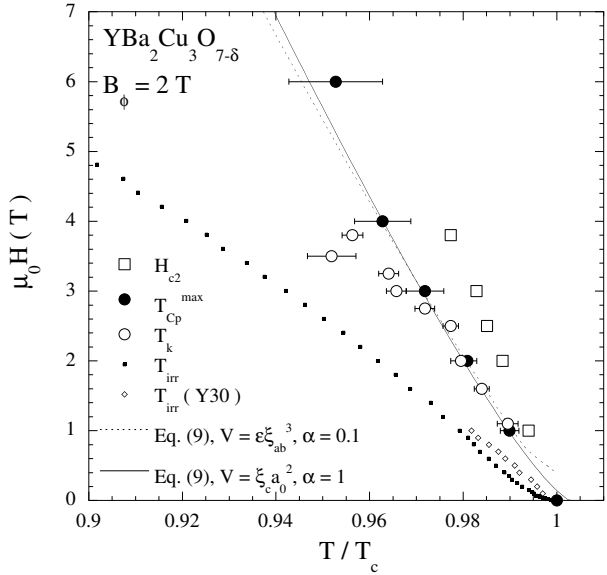}}
  \caption{YBa$_{2}$Cu$_{3}$O$_{7-\delta}$: Comparison of the
   $H_{c2}(T)$ line ( 
  \protect\raisebox{1pt}{\protect\framebox(6,6)[t]{}} \color{black}
  ), $T_{Cp}^{max}(H)$  (\Large $\bullet$\normalsize ),
     $T_{k}(H)$   (\Large $\circ$ \normalsize ), and
     $T_{irr}(H)$ ($\bullet$, 
     \small $\Diamond$ \color{black}
  \normalsize ) for the two crystals with $n_{d} = 1 \times
  10^{11}$ tracks cm$^{-2}$ ($B_{\phi} = 2 $ T). The drawn line
  denotes the locus of $|F_{n}-F_{s}(B)+n_{t}U_{p}(B)| =
  \alpha k_{B}T/V$, with $V = \varepsilon a_{0}^{2}\xi_{ab}$ and 
$\alpha = 1$ (see
  text). The dotted line shows the same, but choosing $V = \varepsilon
  \xi_{ab}^{3}$ and $\alpha = 0.1$.
  }
\label{comp-IRL}
\end{figure}

\noindent with $\beta_{A} = 1.16$. According to Eq.~(\ref{eq:abrikosov}), the zero 
intercepts of $H^{-1}d\Gamma/d\theta$ correspond to $T_{c}(H)$, {\em i.e.} 
the $B_{c2}(T)$ line. A plot of the zero intercepts for different fields yields 
the slope $dB_{c2}/dT = -2$ TK$^{-1}$ and $\xi_{ab}(0) = 1.37$ nm, 
while dividing out the field dependence $1-B/B_{c2}(T)$ gives the line energy 
$\varepsilon_{0}(T)$. The inset (b) of Fig.~\ref{Hc2} shows that $\varepsilon_{0}(T)$ 
does not depend on field, as it should. The results were checked by plotting the
scaled magnetization, $H^{-1}d\Gamma/d\theta / (tb)^{2/3}$ as function of the 
LLL parameter $Q$, see inset (c) of Fig.~\ref{Hc2}. The magnetization 
could be scaled using the experimental $dB_{c2}/dT = -2$ 
TK$^{-1}$ and the mean-field critical temperature $T_{c}^{MF}
 = 92.3 $ K. The linear portion of the $H^{-1}d\Gamma/d\theta$ curves 
falls in the regime $Q > 10$, well outside the  fluctuation-dominated region.
The $B_{c2}(T)$--line, plotted in Fig.~\ref{comp-IRL}, the $\xi_{ab}$ value deduced from it, and 
$d\varepsilon_{0}/dT$, which yields the penetration depth extrapolated to zero temperature 
$\lambda_{ab}(0) = 120$ nm, agree well with the results in the 
literature.\cite{sonier2000}

\begin{figure}[t]
\centerline{\epsfxsize 8cm \epsfbox{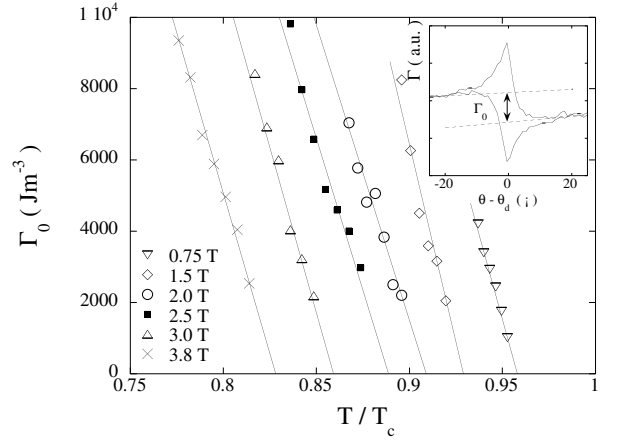}}
  \caption{ K$_{0.35}$Ba$_{0.65}$BiO$_{3}$: Amplitude of the torque
  discontinuity  $\Gamma_{0}$ for field alignment with the defect
  direction $\theta_{d}$. Straight lines show the extrapolation of the torque step
  to $T_{k}(H)$. The inset shows the torque curves for both
  directions of rotation  and the determination of the torque jump as the
  difference between the linear extrapolations of the torque
  from high angle to $\theta = 0$ (dotted lines). The applied field 
   $\mu_{0}H= 1.25$ T, the temperature $T = 0.92 T_{c}$. }
  \label{stepbk}
\end{figure}

As first shown in Ref.~\onlinecite{hayani2000}, and reproduced in the
Inset of Fig.~\ref{step}, the equilibrium torque
features a step of magnitude $\Gamma_{0}$ at the angle at which the field
orientation coincides with the defect direction $\theta = 
\theta_{d}$. On both sides of the step, the torque signal depends linearly on the field angle, 
allowing $\Gamma_{0}$ to be determined from the extrapolation of the torque to $\theta -
\theta_{d} = 0$ (Inset to Fig.~\ref{step}). The main panel of Fig.~\ref{step}
shows $\Gamma_{0}$  as function of  temperature, for fields $0.5 B_{\phi} <
\mu_{0}H < 2 B_{\phi}$. The torque step decreases approximately
linearly with temperature, with a roughly field-independent 
slope $d\Gamma_{0}/dT = 65 \pm 5 $ Jm$^{-3}$K$^{-1}$, and
vanishes at the onset temperature $T_{k}(H)$
[or, conversely, at the field $H_{k}(T)$]. The locus of $T_{k}(H)$, as well
as the irreversibility line $T_{irr}(H)$ and the $B_{c2}(T)$-line, are displayed in Fig.~\ref{comp-IRL}.
Additional measurements on samples with $B_{\phi} = 1$ T and $B_{\phi} = 0.4$ T
indicate that $T_{k}(H)$ depends weakly on the irradiation dose in this
range of $B_{\phi}$.

\begin{figure}[t]
\centerline{\epsfxsize 8cm \epsfbox{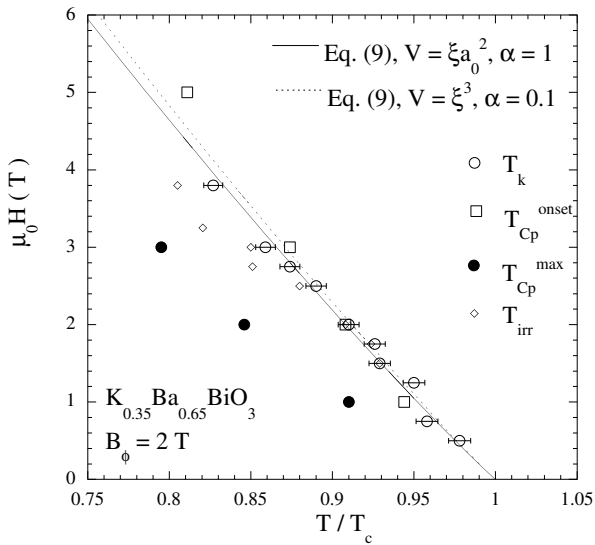}}
  \caption{K$_{0.35}$Ba$_{0.65}$BiO$_{3}$, $B_{\phi} = 2 $ T: Comparison of the
  specific anomaly heat onset temperature $T_{Cp}^{onset}(H)$ ( 
  \protect\raisebox{1pt}{\protect\framebox(6,6)[t]{}} \color{black}
  ), the specific heat maximum $T_{Cp}^{max}(H)$  (\Large 
$\bullet$\normalsize ),
     $T_{k}(H)$   (\Large $\circ$ \normalsize ), and
     $T_{irr}(H)$ ( \small $\Diamond$ \color{black}
  \normalsize ) . The drawn line
  denotes the locus of $|F_{s}(B)+n_{t}U_{p}(B)-F_{n}| =
  \alpha k_{B}T/V$, with $V = a_{0}^{2}\xi$ and $\alpha = 1$ (see
  text). The dotted line shows the same, but choosing $V = \xi^{3}$ 
and $\alpha = 0.1$.
  }
\label{BKBO-comp-IRL}
\end{figure}

Concerning K$_{0.35}$Ba$_{0.65}$BiO$_{3}$, a discontinuity in the 
reversible torque signal,
superimposed to some irreversibility along the track direction,
can be put into evidence in a way similar to YBa$_{2}$Cu$_{3}$O$_{7-\delta}$.
The torque step again depends roughly linearly on temperature, with slope
$d\Gamma_{0}/dT \approx 6 \pm 0.5 \times 10^{2}$ Jm$^{-3}$K$^{-1}$ 
(Fig.~\ref{stepbk}).
The $T_{k}(H)$ line is found to lie slightly above $T_{irr}(H)$
(Figure~\ref{BKBO-comp-IRL}).

For both compounds, the temperature- and field dependence of $\Gamma_{0}$
is found to be parameterized, at all but the lowest measuring fields 
($\mu_{0}H = 1.1$ T for Y30, 0.75 T for K$_{0.35}$Ba$_{0.65}$BiO$_{3}$),
by the LLL parameter $Q$ [see Fig.~\ref{fig:LLL} (a,b)].  All torque step
data trace the same curve when divided by the free energy density
$k_{B}TB/\Phi_{0}\xi(0)$ and plotted vs $Q$, assuming the  mean field
transition temperature to be $T_{c}^{MF} = 93.1$ K in
YBa$_{2}$Cu$_{3}$O$_{7-\delta}$, and 32.4 K in K$_{0.35}$Ba$_{0.65}$BiO$_{3}$.
The condition for thermodynamic quantities to depend on the sole 
parameter $Q$ is that the Landau level splitting $\sim 2 T_{c}^{MF} h$ be greater than the
strength of fluctuations $T_{c}^{MF} (2 Gi)^{1/3} (ht)^{2/3}$ [with
$h \equiv 2 \pi \xi^{2}(0) B/\Phi_{0} = B / T_{c}^{MF}/(\partial 
B_{c2} / \partial T)_{T=T_{c}^{MF}} = (\partial b^{-1}/\partial t)_{t=1}^{-1}$].\cite{Ikeda95} In
YBa$_{2}$Cu$_{3}$O$_{7-\delta}$, at the temperatures under
investigation, this condition is satisfied for fields larger than 1 T.
The failure of the low--field data to conform to the scaling might thus
be due to Landau level degeneracy. In K$_{0.35}$Ba$_{0.65}$BiO$_{3}$,
however, the LLL condition is supposed to be satisfied at all
fields investigated. The fact that the low-field data do not
conform to the LLL-scaling in this compound either, suggest another
origin of the breakdown of scaling. Namely, the intrinsic
inhomogeneity introduced by the randomly positioned columnar defects leads
to a spread in local field-dependent critical temperatures to which the 
superconductor is sensitive at fields $B \lesssim B_{\phi}$.\cite{vdBeek96,Braverman2002} 
As a result, the effective $T_{c}$ is not the real critical temperature but an
average quantity (with a lower value) determined by the defect distribution. 
This effect was previously put into evidence in heavy-ion irradiated 
Bi$_{2}$Sr$_{2}$CaCu$_{2}$O$_{8+\delta}$: at fields $0.2 B_{\phi} < 
B < B_{\phi}$ the magnetization was found to approximately follow
the LLL scaling relation but with an effective critical temperature 2 
K lower than the mean-field $T_{c}$ of the pristine sample. We 
suggest that a similar effect is responsible for the downward shift of the 
present 1 T data in Y30, and the 0.75 T data on the 
K$_{0.35}$Ba$_{0.65}$BiO$_{3}$ crystal, as well as for the difference 
in $T_{c}^{MF}$ deduced for Y30 for field along the tracks and field 
along the $c$-axis (in the latter case, the field component along the defects is 
smaller).


\begin{figure}[t]
     \centering
     \centerline{\epsfxsize 8cm \epsfbox{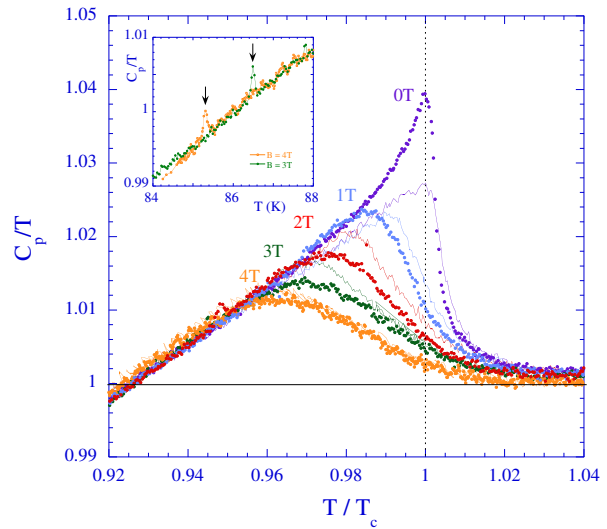}}
     \caption{ YBa$_{2}$Cu$_{3}$O$_{7-\delta}$: Normalized
     specific heat data on a pristine crystal (points), and on a
     crystal with $n_{d} = 1\times 10^{11}$ cm$^{-2}$, \em i.e. \rm
     $B_{\phi} = 2$ T (thin lines). In order to compare the two data sets,
     the temperature has been rescaled to $t = T/T_{c}$, and the
     specific heat to the normal state (phonon) contribution at
     $T_{c}$. The temperature dependence due to phonons was subtracted.
     The inset shows an enlargement of the specific heat curve measured on the
     pristine crystal in fields of 3 and 4 T. }
     \label{fig:YBCO-Cp}
\end{figure}

\subsection{Specific heat}
\label{section:Cp}

Specific heat measurements have been performed on all 
YBa$_{2}$Cu$_{3}$O$_{7-\delta}$
crystals (see table~\ref{tbl:crystals}). The same measurement technique was
employed as in Ref.~\onlinecite{Marcenat2003}.
Well-defined specific heat anomalies were observed in all
crystals, in spite of the use of large irradiation doses (up to $B_{\phi} = 5$ T).

In the pristine
sample,  the amplitude of the zero field superconducting jump was of 
the order of 4\% of the
total specific heat, attesting to its very high quality. The anomaly presents
the typical shape of the superconducting transition in presence of 
strong thermal
fluctuations for $H = 0$, broadens for increasing $H$ and is shifted 
towards lower
temperature (see Fig.~\ref{fig:YBCO-Cp}). Sharp vortex lattice melting peaks
are observed for 1 T$ <H<$ 6 T (see inset of 
Fig.~\ref{fig:YBCO-Cp}).\cite{Schilling96,Bouquet2001}

\begin{figure}[t]
     \centerline{\epsfxsize 8.5cm \epsfbox{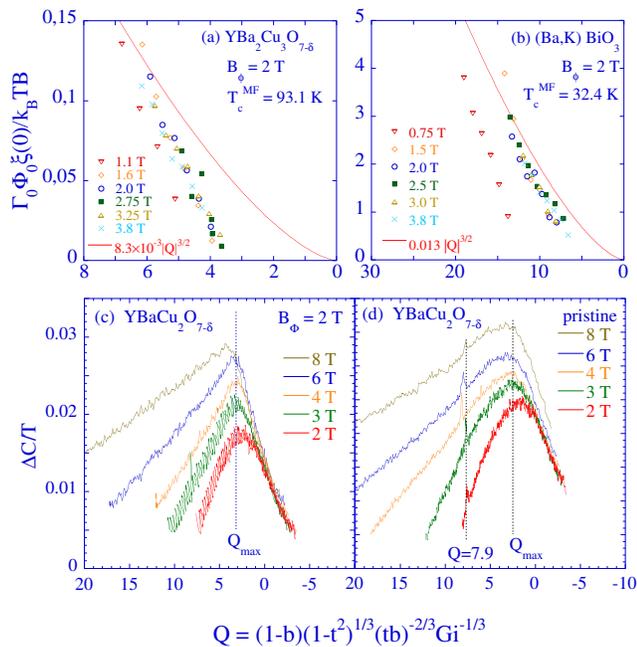}}
      \caption{Lowest Landau level scaling of the torque (a,b) and
      specific heat data (c-f). (a) Scaled torque jump, measured on
      YBa$_{2}$Cu$_{3}$O$_{7-\delta}$ crystal Y30, versus LLL
      parameter $Q$ (see text). The parameter $T_{c}^{MF} = 93.1$ K.
      (b) Scaled torque jump, measured on the
      Ba$_{0.65}$K$_{0.35}$BiO$_{3}$ crystal with $B_{\phi} = 2$ T, versus
      $Q$, with $T_{c}^{MF} = 32.4$ K.
      (c) $C/T$ versus $Q$ for irradiated 
YBa$_{2}$Cu$_{3}$O$_{7-\delta}$ with $B_{\phi} = 2$ T.
      (d) $C/T$ versus $Q$ for pristine YBa$_{2}$Cu$_{3}$O$_{7-\delta}$. }
    \label{fig:LLL}
    \end{figure}

The presence of the amorphous columnar
defects reduces the absolute temperature $T_{C_{p}}^{max}$
at which the specific heat maximum occurs in \em zero \rm
field (see table~\ref{tbl:crystals}).\cite{zerofieldpaper} A lowering of the critical temperature
after heavy ion irradiation may occur as a result of ``self-doping''
of the intercolumn space by O ions expelled from the 
tracks,\cite{Pomar2000,MingLi2002}
but no such effect was reported for YBa$_{2}$Cu$_{3}$O$_{7-\delta}$. Rather,
the columns may act by reducing the average \em zero \rm field $T_{c}$ at
which long range superconducting order can set in.\cite{vdBeek96,Braverman2002}

Turning to the specific heat data in non-zero field, we find that the
columnar defects \em increase \rm the (reduced)
temperature $T_{C_{p}}^{max}(H)/T_{c}$ of the $C_{p}$ maximum 
(Fig.~\ref{fig:YBCO-Cp}).
Furthermore, the specific heat curves in magnetic field become 
sharper after irradiation. Apparently, columnar defects suppress 
order parameter fluctuations in a magnetic field,
even though the effect in  YBa$_{2}$Cu$_{3}$O$_{7-\delta}$ is weaker 
than that previously found in K$_{x}$Ba$_{1-x}$BiO$_{3}$.\cite{Marcenat2003,Klein2004}  
No vortex lattice melting anomaly is observed in the irradiated crystals, nor
is any other anomaly at, {\em e.g.}, the irreversibility line.
    
Figure~\ref{comp-IRL} shows that in YBa$_{2}$Cu$_{3}$O$_{7-\delta}$,
the maximum of the specific heat systematically coincides with the 
temperature $T_{k}$ at which the torque signal from vortex pinning 
by the columnar defects disappears.
Thus, {\em vortex pinning by the columns is responsible for the upward shift of
the superconducting transition}. This is the obvious when
one considers the shift for different columnar defect densities
$n_{d}$ (matching fields $B_{\phi}$). Figure~\ref{IRL-Cp} shows that a higher
density of columns leads to a higher $T_{C_{p}}^{max}$.

Figure~\ref{BKBO-comp-IRL} traces the {\em onset} temperature 
$T_{C_{p}}^{onset}(H)$ of the specific heat anomaly measured in 
K$_{0.35}$Ba$_{0.65}$BiO$_{3}$,\cite{Marcenat2003,Klein2004}
along with the $T_{k}(H)$-- and $T_{irr}(H)$ lines obtained here. In
the bismuthate compound the two latter lines coincide with the specific
heat {\em onset}, which is much sharper than the one in 
YBa$_{2}$Cu$_{3}$O$_{7-\delta}$.
This attests to the fact that in this compound, the effect of the columnar
defects is important enough to not only remain present up to
$H_{c2}(T)$, but to actually determine the position of the upper
critical field line (see also Refs.~\onlinecite{Marcenat2003,Klein2004}).

In the LLL scenario,  the specific heat in magnetic 
field should behave as $C_{p}/T = F(Q)$, with $F$ a universal scaling 
function.\cite{Welp91,DingpingLi2001}
Figure~\ref{fig:LLL}(c-d) shows that the location in the $(H,T)$ 
plane of characteristic features of $C_{p}$ are satisfactorily described by $Q = constant$.
For example, vortex lattice melting in the pristine crystal
occurs at constant $Q = 7.9$,\cite{DingpingLi2002} and the position
of the maximum at $Q = 3.2$. However, as in Refs.
\onlinecite{Welp91} and \onlinecite{Junod94}, the magnitude of $\Delta C_{p}/T = C_p-C_N$ does
not scale as expected, whatever the choice for the normal contribution ($C_N$) (see Figure~\ref{fig:LLL}(c-d)).

\section{Data analysis and Discussion}

\subsection{Determination of the pinning energy}
\label{section:pinning}

The torque exerted by a perpendicular magnetic field component on the 
trapped vortices
allows one to measure the energy gain due to vortex localization on 
the columnar
defects,\cite{Drost98} {\em i.e.} the pinning energy.
Writing out the
torque step $\Gamma_{0} \simeq H_{\perp}M_{\parallel} - H_{\parallel}
M_{\perp} |_{\theta = \theta_{d}} = - H_{\parallel} \partial G / \partial B_{\perp}$
shows that it is a measurement of the energy per unit length
$E_{k} =  \Phi_{0} \partial G / \partial B_{\perp}$ of vortex kinks 
joining different defects ($G$ is the Gibbs
free energy). In the limit of \em isolated \rm flux lines,
$\Gamma_{0} \approx 2 E_{k} / a_{0}^{2}$, where $a_{0} \simeq
(\Phi_{0}/B)^{1/2}$ is the flux line spacing. The kink energy
$E_{k} = (\frac{1}{2}U_{p}\varepsilon_{l})^{1/2}$, where $U_{p}$
is the pinning energy of the defect per unit length, and
$\varepsilon_{l}$ is the vortex line tension.\cite{Blatter94}
For a single vortex line, the pinning energy can be written as\cite{Blatter94}

\begin{eqnarray}
    U_{p} & = & \frac{k_{B}^{2}T^{2}}{4 \varepsilon_{l}\xi^{2}}  \beta
        \hspace{3.4cm} (\beta \gg 1 ) \label{eq:pinning-a} \\
     U_{p} & = & \frac{k_{B}^{2}T^{2}}{4 \varepsilon_{l}\xi^{2}}  \beta
     \exp \left( -\pi/\sqrt{\beta}\right) \hspace{1cm} (\beta \ll 1).
     \label{eq:pinning}
\end{eqnarray}

\begin{figure}[t]
\centerline{\epsfxsize 8.5cm \epsfbox{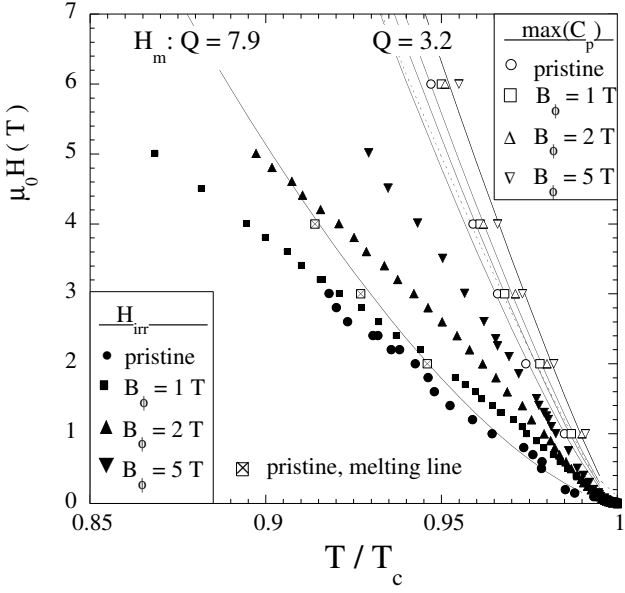}}
  \caption{ YBa$_{2}$Cu$_{3}$O$_{7-\delta}$: Closed symbols denote
  the irreversibility lines $H_{irr}(T)$ determined from the onset of 
the third harmonic
  transmittivity, for  single crystals with $B_{\phi}=0$ 
(\color{black} $\bullet$
  \color{black} ),
  $B_{\phi}=1$ T (\color{black}\protect\rule{1,5mm}{1,5mm}\color{black}),
  $B_{\phi}=2$ T ( \color{black} $\triangle$ \color{black} )
  and $B_{\phi}=5 $ T (\color{black}$\nabla$\color{black}). Open symbols
  (\color{black}$\circ$\color{black}, \color{black}$\Box$\color{black},
  \color{black}$\triangle$\color{black},
  \color{black}$\nabla$\color{black}) correspond to the
  $T_{C_{p}}^{max}$--line tracing the temperature of the specific heat maximum,
  for the same crystals. The drawn lines denote the solution of
  Eq.~(\protect\ref{eq:3point2}), for matching fields $B_{\phi} = 0, 
1, 2$, and 5 T (see text).
  The dotted line shows an evaluation of criterion (\protect\ref{eq:Cooper}),
  choosing  $V = \varepsilon  \xi_{ab}^{3}$ and $\alpha = 0.1$, for
  $B_{\phi} = 1$ T only. The crossed squares depict the vortex lattice melting
  line of the pristine crystal, which is well described by $Q = 7.9$.
  }
\label{IRL-Cp}
\end{figure}

\noindent The pinning strength $\beta$ can be approximated as
$\beta_{core} = 
(c_{0}^{2}\varepsilon_{0}\varepsilon_{l}/k_{B}^{2}T^{2})(1-b)^{2}$
for vortex core pinning (for $c_{0} \ll \xi$) and
$\beta = \beta_{em} = 
(\xi^{2}\varepsilon_{0}\varepsilon_{l}/k_{B}^{2}T^{2})(1-b)$
for electromagnetic pinning ($c_{0} \gg \xi$).\cite{Blatter94} The 
exponential factor
in Eq.~(\ref{eq:pinning}) expresses the reduction of the pinning
energy due to vortex line wandering when $\beta \ll 1$. As a
consequence, the torque jump

\begin{eqnarray}
    \Gamma_{0} & = & \sqrt{2} \frac{k_{B}T}{a_{0}^{2}\xi} \beta^{1/2}  
    \hspace{32mm} (\beta \gg 1 ) \\
     \Gamma_{0} & = & \sqrt{2} \frac{k_{B}T}{a_{0}^{2}\xi} \beta^{1/2} \exp
     \left( - \pi/\beta^{1/2} \right)  \hspace{7mm}  (\beta \ll 1 )
     \label{eq:torquejump-hiT}
\end{eqnarray}

\noindent is a direct measure of the pinning strength $\beta$. Note
that in the \em single vortex limit \rm, the dimensionless torque plotted in
Fig.~\ref{fig:LLL}(a,b) would directly correspond to the square-root 
of the pinning
strength $\beta$.

In practice though, experiments are rarely carried out in the isolated
vortex limit. At high fields, only a fraction $N_{t} = 1 -
\exp(-a_{0}^{2}n_{d}) \approx B_{\phi}/B $ of the
vortices are trapped by the columns. Then we can write

\begin{equation}
     \Gamma_{0} = \sqrt{2} \frac{k_{B}TB_{\phi}}{\Phi_{0}\xi} \beta^{1/2}
     \hspace{1cm} (\beta \gg 1) ;
     \label{eq:torque-hiB-loT}
\end{equation}

\noindent taking $\beta = \beta_{core}$, we can rewrite this as the ``mean--field 
expression''

\begin{equation}
     \frac{\Phi_{0}\xi(0)}{k_{B}TB}\Gamma_{0} = \frac{1}{\sqrt{2}}
     \frac{c_{0}}{2\pi\xi(0)} \frac{B_{\phi}}{B_{c2}(0)} |Q|^{3/2} 
\hspace{1cm} (\beta \gg 1) .
     \label{eq:torque-LLL}
\end{equation}

\noindent In other words, in high fields the torque jump normalized by the energy
density scale $k_{B}T/a_{0}^{2}\xi(0)$ is expected to follow an 
expression that only depends on the LLL scaling parameter $Q$. Note 
that the functional dependence on temperature and field following the 
LLL parameter is only followed 
\em provided vortex line wandering is not important \rm ($\beta \gg 1$). 
The exponential factor in Eq.~(\ref{eq:torquejump-hiT}) 
related to vortex line wandering would spoil the scaling, 
because its argument cannot be expressed as a function of $Q$.
The irrelevance of line wandering is slightly surprising, for our 
experiments are carried out at high temperatures at which it
is expected to be relevant.\cite{KrusinElbaum94,Samoilov96,Blatter2002} However,
experiment unambiguously shows that this is not the case. In other words, 
a description of pinning in terms of vortex line fluctuations only (\em i.e. \rm the London model)
is \em not \rm a good starting point for the description of vortex physics in
heavy-ion irradiated high temperature superconductors in Tesla
fields, regardless of their anisotropy. The reason is that vortex 
line fluctuations can be interpreted as the result of superposing thermally generated 
vortex loops on regular, field-generated vortices.\cite{nguyen99} At high fields at which the
LLL condition $4  B \xi^{2}(0)/\Phi_{0} >  Gi$ is satisfied, thermal
vortices cannot be excited, as these imply Landau-level
degeneracy.\cite{Ikeda95} Rather, our data show that the fluctuation of the
order parameter amplitude must be considered when describing
pinning in the vortex liquid, as was already suggested in
Ref.~\onlinecite{Ikeda96}.

The importance of order parameter amplitude fluctuations to pinning becomes
apparent when we compare expression (\ref{eq:torque-LLL}) to the
experimental data, see Fig.~\ref{fig:LLL}(a,b). For
YBa$_{2}$Cu$_{3}$O$_{7-\delta}$, the
experimental torque lies well below the prediction (\ref{eq:torque-LLL}).
The experimental data are consistent with either an exponential,
$(a_{0}^{2}\xi(0) /  k_{B}T) \Gamma_{0} \propto e^{-Q}$, or with a
power--law drop, $(a_{0}^{2}\xi(0) /  k_{B}T) \Gamma_{0} \propto
Q^{4}$, or with a polynomial in $Q$, but shows no sign of divergent behavior
on approaching the irreversibility line.\cite{Tesanovic94} In
K$_{0.35}$Ba$_{0.65}$BiO$_{3}$ however, the experimental pinning
energy lies close to the predicted value (\ref{eq:torque-LLL}).
Fluctuations have little importance here due to the small Ginzburg number
of K$_{0.35}$Ba$_{0.65}$BiO$_{3}$.

\subsection{Shift of the specific heat maximum}
\label{section:suppression}

We finish by showing that vortex core pinning by the columnar
defects, such as described by Eq.~\ref{eq:torque-LLL}, leads to a
correct estimate of the specific heat shift in both
YBa$_{2}$Cu$_{3}$O$_{7-\delta}$ and K$_{0.35}$Ba$_{0.65}$BiO$_{3}$,
and may thus account for the free energy gain obtained from pinning.
We adopt the procedure of Ref.~\onlinecite{Klein2004}, in which the
field-angle dependence and the defect density-dependent shift of the 
onset temperature
$T_{C_{p}}^{onset}(H)$ in K$_{x}$Ba$_{1-x}$BiO$_{3}$ was well described
\em quantitatively \rm . To estimate 
$T_{C_{p}}^{onset}(H)$, we add the
free energy change $n_{t}U_{p}(B)$ from pinning to the free energy difference
$F_{n} - F_{s}(B) = \frac{1}{2}\mu_{0}H_{c2}^{2}(1-b)^{2}$
of the normal and superconducting states of the pristine
material, and equate \cite{Cooper95,Klein2004}
\begin{equation}
\left|F_{n}-F_{s}(B)+n_{t}U_{p}(B)\right| = \alpha \frac{k_{B}T}{V}.
\label{eq:Cooper}
\end{equation}
Here $n_{t} = N_{t}a_{0}^{-2}$ is the areal density of vortices 
trapped on a columnar
defect, $U_{p}$ is the average pinning energy per vortex per unit
length, given by Eq.~(\ref{eq:pinning-a}), $V = \varepsilon\xi^{3}$ is 
the coherence volume,
and $\alpha \lesssim 1$. In Ref.~\onlinecite{Klein2004}, Eq.~(\ref{eq:Cooper})
was found to quantitatively reproduce all lines [$T_{irr}(H), T_{k}(H),
T_{C_{p}}^{onset}(H)$] if one chooses $U_{p}$ to be given by the expression for
electromagnetic pinning, Eq.~(\ref{eq:pinning}) with $\beta =
\beta_{em}$ and $\alpha \approx 1$. There is some latitude in the choice of
parameters: the dotted line in Figure~\ref{BKBO-comp-IRL} shows
that similar good agreement can be obtained with $\beta =
\beta_{core}$ and $\alpha = 0.1$.

We find that the position of the $T_{C_{p}}^{max}$--line in 
YBa$_{2}$Cu$_{3}$O$_{7-\delta}$
can also be described in this way. The dotted line in Fig.~\ref{comp-IRL} shows
the solution of Eq.~\ref{eq:Cooper} with parameters for
YBa$_{2}$Cu$_{3}$O$_{7-\delta}$, the pinning energy as given by
Eq.~(\ref{eq:pinning}) with $\beta = \beta_{core}$, and $\alpha = 0.1$.

Another approach is to choose the correlation volume
$V = \varepsilon a_{0}^{2}\xi$ rather than $\varepsilon \xi^{3}$. In
the absence of columnar defects ($n_{t} = 0$), Eq.~(\ref{eq:Cooper}) 
can then be
reduced to
\begin{equation}
     Q^{3/2} = 4 \pi \sqrt{2} \alpha,
     \label{eq:melting-line}
    \end{equation}
\noindent which, for $\alpha = 1.25$ ($Q = 7.9$),  perfectly describes the
position of the vortex lattice melting line (see Fig.~\ref{IRL-Cp}).
In the presence of columnar defects, one again adds the free energy 
gain $n_{t}U_{p}$.
Taking the same expressions $U_{p} =(c_{0}/2\xi )^{2} 
\varepsilon_{0}(1-b)^{2}$ and
$n_{t} = B_{\phi}/\Phi_{0}$  used to derive Eq.~(\ref{eq:torque-LLL}),
Eq.~(\ref{eq:Cooper}) becomes
\begin{equation}
     Q^{3/2} = \frac{4 \pi \sqrt{2} \alpha}{1 + B_{\phi}/B_{c2}(0)}.
     \label{eq:Cppeak}
\end{equation}
\noindent In other words, the criterion (\ref{eq:Cooper}) takes the form
``$Q = constant$'', where the constant depends on the density
of columnar defects. In order to describe the shift of the specific
heat maximum with $B_{\phi}$, we adopt the critical temperature $T_{c}^{MF} =
93.1$ K obtained from the scaling of the torque and the specific heat,
and evaluate
\begin{equation}
     Q = 3.2 / [ 1 + B_{\phi}/B_{c2}(0) ]^{2/3}
     \label{eq:3point2}
     \end{equation}
so as to recover the correct position of the specific heat maximum
for $B_{\phi} = 0$ (see Fig.~\ref{fig:LLL}). The dependence (\ref{eq:3point2})
well describes the upwards shift of the specific heat
maximum in a magnetic field (Fig.~\ref{IRL-Cp}). Thus, we find that
the incorporation of the pinning energy in the free energy difference
between the normal and superconducting states satisfactorily describes
the evolution of the specific heat as function of defect density.

\section{Summary and conclusions}

Thermodynamic measurements in the vortex liquid of
heavy-ion irradiated YBa$_{2}$Cu$_{3}$O$_{7-\delta}$ show that, as in
K$_{x}$Ba$_{1-x}$BiO$_{3}$, the reduction of the average free energy of
the superconductor due to vortex pinning on columnar defects remains
important all the way into the regime of strong order parameter
amplitude fluctuations. Notably, the effect of pinning on the free
energy is sufficient to shift the superconducting transition, as
measured by the heat capacity, further upwards as the defect density
increases. The scaling of the pinning energy with the LLL parameter
$Q$ shows that pinning is affected by fluctuations of the overall
order parameter amplitude. These should therefore be taken into
account in any description of pinning in the vortex liquid - a model
based on vortex line positional fluctuations only is bound to be
inadequate. However, due to the lower Ginzburg number, order
parameter fluctuations are much less important in the bismuthate
superconductor than in YBa$_{2}$Cu$_{3}$O$_{7-\delta}$.

This brings us to the final point, which is the Bose-glass transition.
Our results indicate that depinning of vortices from columnar defects
due to line wandering is not likely to be a very effective mechanism
for this transition. On the other hand, fluctuations of the overall
order parameter amplitude are also expected to become
unimportant at, or near to, the Bose-glass transition. We note that, 
contrary to
the melting line in the pristine crystals, the irreversibility line
(and especially its high--field part) {\em cannot} be described as a line of
constant $Q$ and therefore does not satisfy LLL scaling. We therefore
suggest that another delocalization mechanism drives the
Bose-glass transition. This could be, for example, vortex delocalization
due to variations of the pinning potential. Analogously to the electronic
mobility in disordered semiconductors, the most weakly bound vortices
could be delocalized  at a much lower temperature than the bulk of the
vortex matter.\cite{Larkin95} Another possibility is that the plastic 
properties
of the vortex ensemble play a much more important role than hitherto
considered. For example, the trapped vortex ensemble could constitute
a polycristal, vortex delocalization taking place on the grain
boundaries. In both cases, another lower energy scale than the 
average pinning energy is
involved in vortex delocalization at $T_{irr}$, a scale that manifests
itself through the large separation between $T_{irr}$ and
$[T_{k}(H),T_{C_{p}}(H)]$ lines in YBa$_{2}$Cu$_{3}$O$_{7-\delta}$.

The situation in cubic K$_{x}$Ba$_{1-x}$BiO$_{3}$ seems to be
different. We have shown that the disappearance of pinning by
columnar defects in this compound happens indistinguishably close to the
superconducting-to-normal boundary. It means that in this compound,
vortices are neither depinned, nor delocalized at all, and the Bose
glass phase subsists up to the temperature at which
superconductivity disappears altogether.

\end{document}